# Using the cathode surface of straw tube for measuring the track coordinates along the wire

V. A. Baranov[1], B. A. Chekhovskiy[2], N. P. Kravchuk[1], A. S. Korenchenko[1], N. A. Kuchinskiy[1*], N. V. Khomutov[1], S. A. Movchan[1], V. S. Smirnov[1], F. E. Zyazyulya[2]

[1] Joint Institute for Nuclear Research, 6 Joliot-Curie 141980 Dubna, Moscow Region, Russia

[2] The National Scientific and Educational Centre of Particle and High Energy Physics of the Belarusian State University (NC PHEP BSU), 153 M. Bogdanovich Street, 220040 Minsk, Belarus

---

[*] corresponding author, tel. 007-496-21-65173, e-mail: kuchinski@jinr.ru

The coordinate detectors based on straw tubes provide a high accuracy of the radial coordinate measurement using the drift time and a small amount of matter in the way of the measured particles. However, the measurement of the coordinate along the wire constitutes a problem. This paper proposes a method for measuring the hit coordinate along the wire with an accuracy better than 1 mm in a straw tube detector using the signals from the cathodes of the detector.

Tracking detectors using separate thin-walled tubes of small diameter (straw tubes) are now widely used in large experimental facilities for high energy physics [1]. Detectors of this type have a small radiation length, and cylindrical geometry provides high electrostatic and mechanical properties, ability to work under high pressure. It is important that each tube is independent of the other ones, and a failure of one tube (including wire breakage) does not violate operational capability of the detector as a whole.

The radial coordinate in the straw tube is determined by the drift time. To measure the coordinate of the hit along the wire (z-coordinate) the additional external strips are currently used in most cases. This allows one to get a good (up to 100 μm) spatial resolution along the anode wire, but it complicates the design of the detector and contributes additional material [2]. Such a solution was proposed for the Mu2e project to search for the μ → e conversion at FNAL, USA [3].

In this paper we propose a principally new way of measuring coordinates along the wire for the straw tubes based on information received from the tube cathode without adding any extra substance into the detector. The proposed measurement method is a combination of two methods. The first stage of a measurement of the coordinate along the anode wire utilizes the principle of current division [4]. In this way, for the anode wire ohmic resistance of 2200 ohm/m the accuracy of less than 2% of the length of each tube can be achieved what gives a rough estimate of the hit coordinate. For a more accurate determination of the hit location a so-called "double wedge" structure on the cathode surface [5] is used.

For the measurements we used the straw tubes of 10 mm in diameter made of aluminum coated Mylar tape by ultrasonic welding technique developed at LHEP JINR [6]. The desired cathode pattern is etched on the surface of 30 micron thick aluminum coated Mylar tape. The shape of the cathode strip pattern used is shown in Fig. 1.

It forms a cathode surface that is divided electrically into two zigzag strips with a periodic structure and a step of 80 mm. The choice of the cathode pattern step is motivated by an estimated typical tube length of 2000 mm and the accuracy of the method of current division of less than 2%. In this case, the current division accuracy is less than a half-period of the strip pattern.

Since the coordinate along the wire is determined by the ratio of charge induced on the tube strips, the measurement accuracy is determined mainly by the ratio of the circumference of the tube to the half-period of the cathode pattern and, to a lesser extent, mechanical inaccuracies of manufacturing the cathodes and the tube itself, the noise of electronics. In addition, the signals on the cathodes in our detector are larger than ones for a planar geometry with the same anode-cathode spacing, and their ratio does not depend on the anode signal amplitude due to cylindrical symmetry of the cathode relative to the anode wire.

The experimental setup used for measurements and the measurement scheme are shown in Figs. 2 and 3 respectively.

The tube was irradiated with electrons from a $^{90}$Sr source. The measurement was triggered by the signals from the anode wire in coincidence with the scintillation counter signals. After the KATOD-1 amplifier [7] the cathode signals were fed into the CAEN VME V1720 digitizer (12 bit, 250 MHz) [8] connected to a PC.

The results obtained are shown in Fig. 4. The analytically calculated solid curve [9] is superimposed on the experimental points. A good agreement between two kinds of results is observed.

The accuracy of the coordinate measurement along the wire can be estimated from Fig. 5 as less than 1 mm given that the width of a collimated electron beam from the source equals 2 mm.

The method using the cathode surface with two cathode strips (Fig. 1) has an ambiguity in the definition of z-coordinates caused by symmetry of the ratio of the charges induced on the strips with respect to a zigzag pattern turning point. This ambiguity could be eliminated by introducing the third cathode strip or by shifting the tube layers relative to each other since the tubes are always assembled in several layers.

This method of determining the track coordinate along the anode wire works in the case of a single track in a straw tube. Multiple tracks could produce pile-up signals on the cathodes.

As the intensity of the accelerator beams and event multiplicity are growing the higher rate capability of the detectors and the straw tubes in particular becomes a very topical issue. We are currently exploring ways to increase the rate capability of the straw tubes. Results will be published in a forthcoming paper.

The authors thank V. M. Grebenyuk, D. L. Demin for fruitful discussions and V. A. Nikitin for his interest. This work was supported by RFBR grant 11-02-01472-a, and BRFFR grant F10D-006.

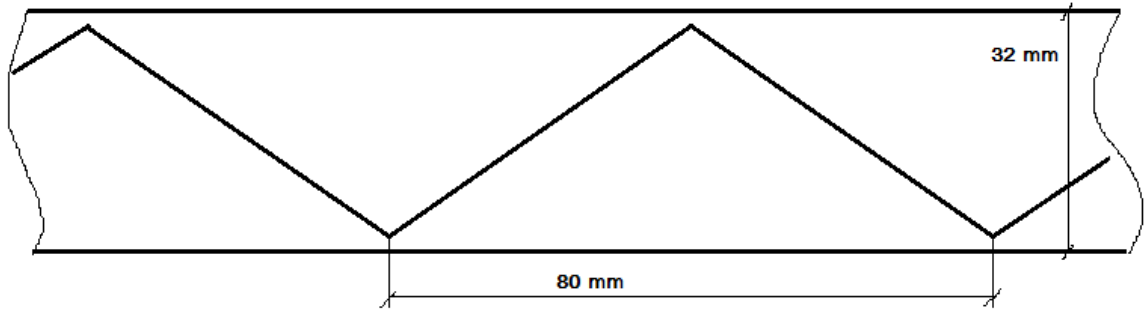

Figure 1: Unfolded surface of the straw tube with two cathodes used in the tests.

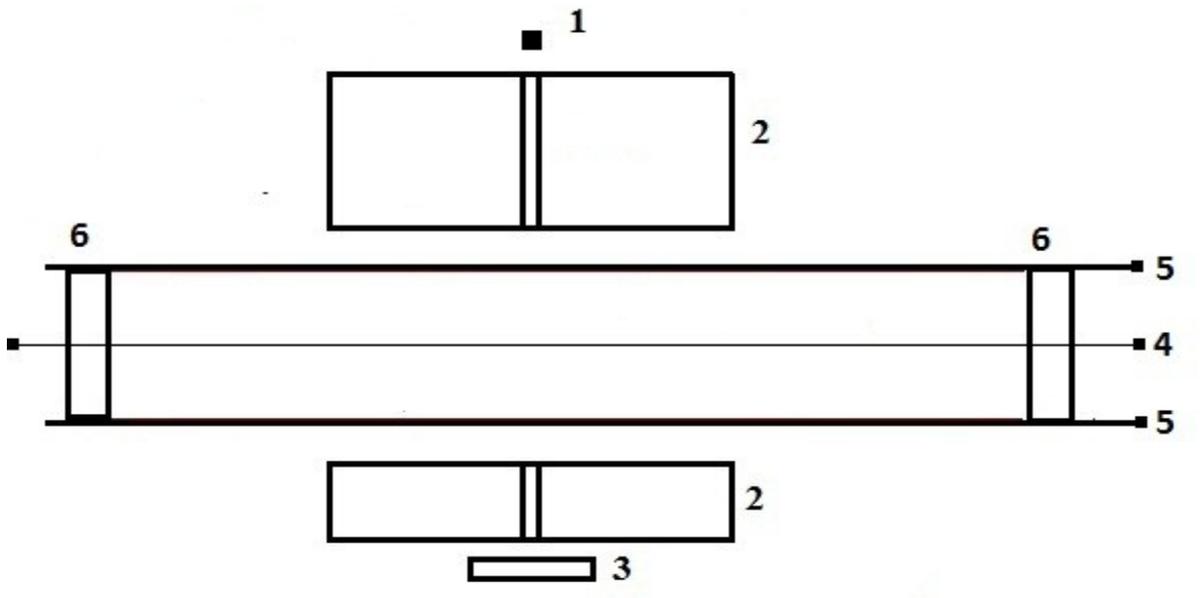

Figure 2: Experimental setup. 1 - $^{90}$Sr source, 2 - collimators with a 2 mm round hole, 3 – 5 mm × 5 mm × 2 mm scintillator, 4 - anode wire output, 5 - cathode outputs, 6 - straw tube end caps.

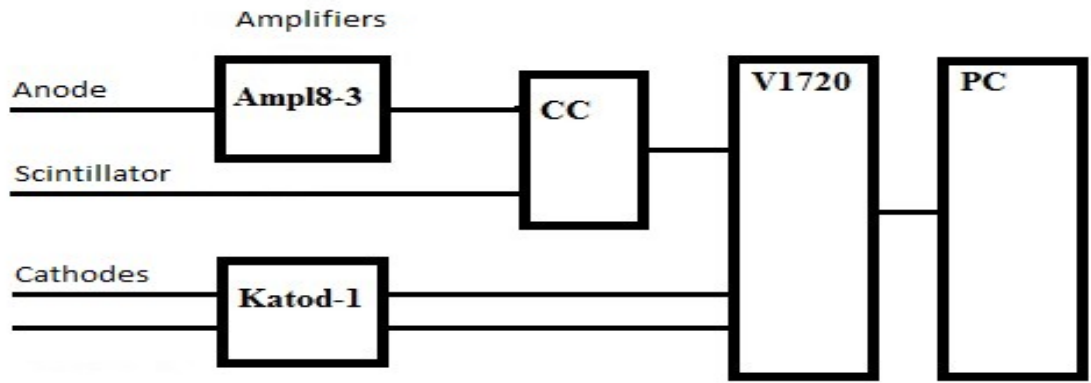

Figure 3: Measurement scheme. CC - coincidence circuit, V1720 - pulse shape digitizer (CAEN, 12 bit, 250 MHz) [8], Katod-1 and Ampl8-3 — Amplifiers.

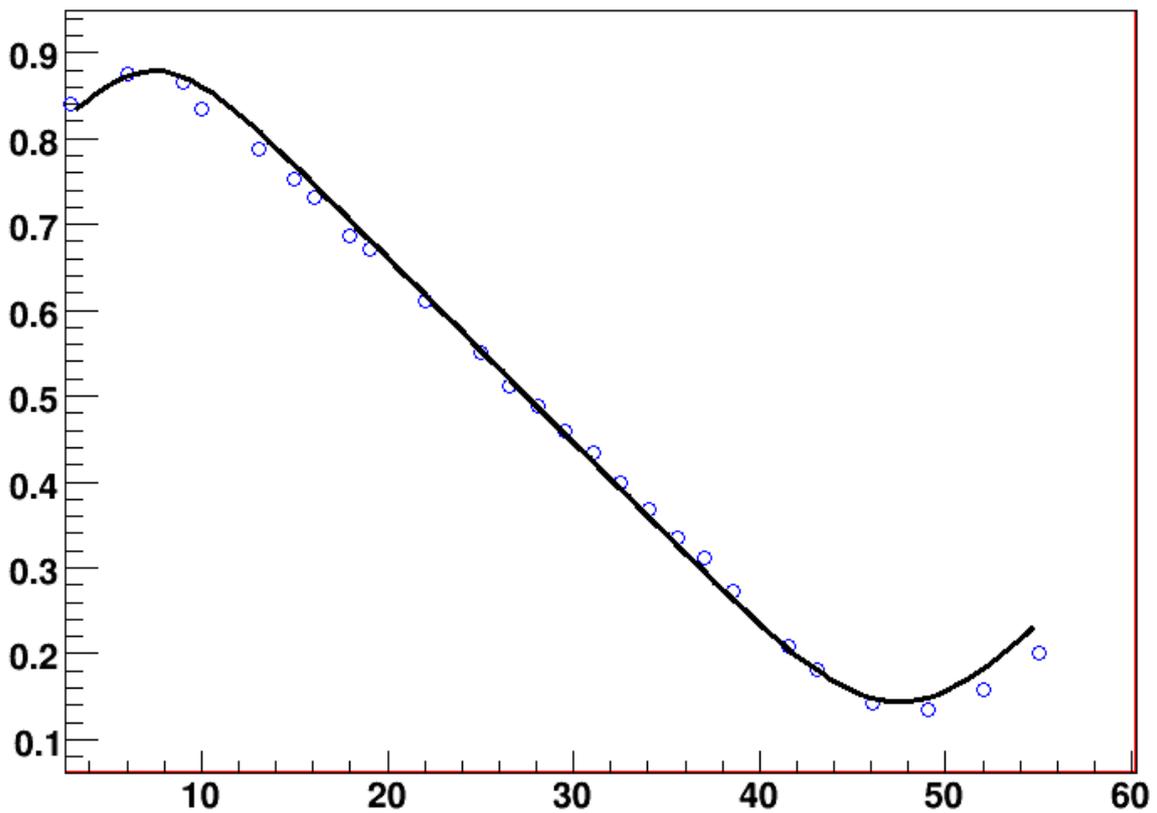

Figure 4: The ratio of the charge induced on one of the cathodes to the total charge as a function of the source position (mm). The solid line is the result of calculations.

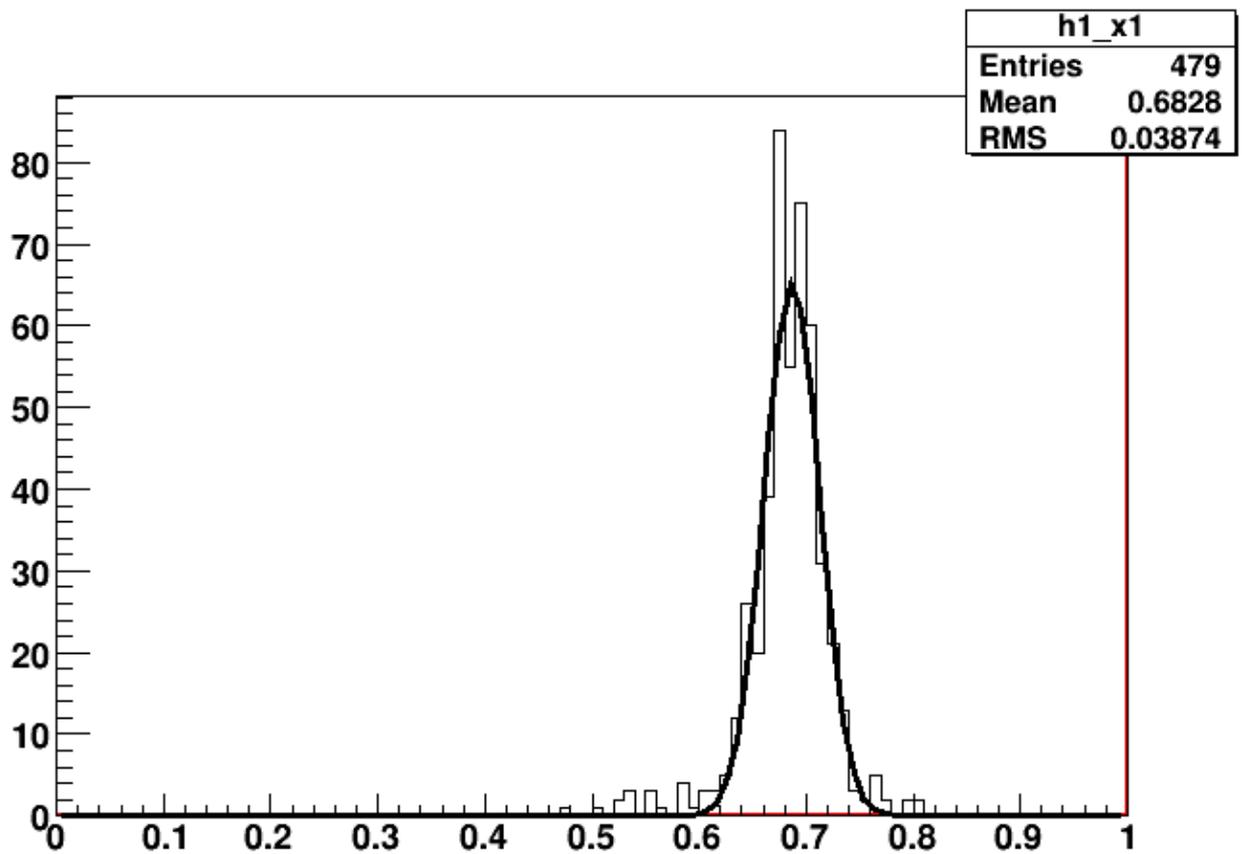

Figure 5: The distribution of the ratio of the charge induced on one of the cathodes to the total charge for a source fixed position along the straw tube.